\documentclass[prb,amsmath,amsfonts,amssymb,nofootinbib,twocolumn]{revtex4}
\usepackage{graphicx}
\usepackage{bm}
\usepackage{color}

\begin{document}
\bibliographystyle{apsrev}
\title{Constrained-Pairing Mean-Field Theory. IV.  Inclusion of corresponding pair constraints and connection to unrestricted Hartree-Fock theory}
\author{Takashi Tsuchimochi}
\affiliation{Department of Chemistry, Rice University, Houston, TX 77005-1892}
\author{Thomas M. Henderson}
\author{Gustavo E. Scuseria\footnote{Corresponding Author: guscus@rice.edu}}
\affiliation{Department of Chemistry and Department of Physics and Astronomy, Rice University, Houston, TX 77005-1892}
\author{Andreas Savin}
\affiliation{Laboratoire de Chimie Th\'eorique, CNRS, Universit\'e Pierre et Marie Curie, 4 Place Jussieu, F-75252 Paris, France}
\date{\today}

\begin{abstract}
Our previously developed Constrained-Pairing Mean-Field Theory (CPMFT) is 
shown to map onto an Unrestricted Hartree-Fock (UHF) type method if one 
imposes a corresponding pair constraint to the correlation problem that forces 
occupation numbers to occur in pairs adding to 1.  In this new version, 
CPMFT has all the advantages of standard independent particle models (orbitals 
and orbital energies, to mention a few), yet unlike UHF, it can dissociate 
polyatomic molecules to the correct ground-state restricted open-shell 
Hartree-Fock atoms or fragments.
\end{abstract}
\maketitle

\section{Introduction}
In a recent series of papers,\cite{CPMFT1,CPMFT2,CPMFT3} we have developed 
constrained-pairing mean-field theory (CPMFT), a method capable of describing 
static (strong) correlation in an accurate and efficient manner.  The idea 
behind CPMFT is to make use of the pairing correlations (see below) that occur 
in a quasiparticle picture to describe static correlation in molecular 
systems. In CPMFT, we divide the natural orbitals into core, active, and 
virtual blocks; each core orbital has unit occupation, each virtual orbital 
has zero occupation, and the active natural orbitals have fractional 
occupations $n_i$, where $0 < n_i < 1$.  Static correlation is introduced 
by allowing electron pairs to have fractional occupations within an active 
space.  

The use of a pairing interaction has many advantages.  Unlike unrestricted 
Hartree-Fock (UHF), CPMFT has zero spin density everywhere for closed-shell systems.  In the absence 
of static correlation, CPMFT reduces to restricted Hartree-Fock (RHF), while 
it dissociates polyatomic molecules to restricted open shell Hartree-Fock 
(ROHF) atoms or fragments.  Essentially, the dissociation limit of CPMFT can 
be thought of as an ensemble solution.  By reducing to RHF in the absence of 
strong correlation and ROHF at dissociation, CPMFT cleanly separates static 
from dynamic correlation, as previously shown in Ref. \onlinecite{CPMFT3}, 
where the CPMFT $\mathbf{P}$ and $\mathbf{K}$ density matrices
were used to construct alternative densities to be used as inputs into 
traditional density functionals for the dynamical correlation energy.  
Remarkably, CPMFT accomplishes these feats at a mean field computational cost 
instead of the combinatorial blowup of complete active space (CASSCF) or full 
configuration interaction (FCI).

While CPMFT is clearly distinct from UHF, it shows some unexpected 
connections.  We can take advantage of these connections to simplify the 
formalism, make it more efficient, and establish interesting similarities.  
The purpose of this paper is to demonstrate these relations and the 
accompanying reformulation of CPMFT. Accordingly, we discuss this connection 
in Sec. \ref{Sec:UHF} at some length, and show how we can use it to simplify 
the solution of the CPMFT equations.  Section \ref{Sec:Results} shows some 
numerical examples, and we provide conclusions in Sec. \ref{Sec:Conclusions}.  
We include an Appendix that discusses some other formal properties of CPMFT.  
First, however, we provide a brief introduction to pairing correlations in 
Sec. \ref{Sec:Pairing}.

\section{Pairing Correlations and the Quasiparticle Picture
\label{Sec:Pairing}}
Strong correlations in nuclear physics or superconductivity are often 
described as the formation of Cooper pairs.  The theoretical machinery which 
does this is the Hartree-Fock-Bogoliubov (HFB) method.\cite{Blaizot}  In HFB, 
we write the wave function $|\Phi_\mathrm{HFB}\rangle$ as a single determinant 
of quasiparticles created by quasiparticle creation operators which are linear 
combinations of electron creation and annihilation operators.  The 
quasiparticle wave function thus violates electron number conservation.  
Because the quasiparticle wave function is a single determinant, its 
associated density matrix $\mathbf{R}$ is idempotent 
($\mathbf{R}^2 = \mathbf{R}$) and Hermitian 
($\mathbf{R} = \mathbf{R}^\dagger$).  We have
\begin{equation}
\mathbf{R} = 
\begin{pmatrix}
\bm{\gamma}  & \bm{\kappa} \\ -\bm{\kappa}^\star & \bm{1} - \bm{\gamma}^\star
\end{pmatrix}.
\end{equation}
Here, $\bm{\gamma}$ is the physical density matrix in the spinorbital basis;  
it is Hermitian but \textit{not} idempotent.  Information about pairing 
correlations is carried by the anomalous density matrix $\bm{\kappa}$, which 
is antisymmetric by definition because 
$\kappa_{ij} = \langle a_i^\dagger a_j^\dagger \rangle$.  We limit our 
discussion to the closed shell case, in which case we have, for 
$\alpha\alpha$, $\alpha\beta$, $\beta\alpha$, and $\beta\beta$ blocks
\begin{subequations}
\begin{align}
\bm{\gamma} &= 
\begin{pmatrix} \mathbf{P} & \bm{0} \\ \bm{0} & \mathbf{P} \end{pmatrix},
\\
\bm{\kappa} &= 
\begin{pmatrix} \bm{0} & \mathbf{K} \\ -\mathbf{K} & \bm{0} \end{pmatrix},
\end{align}
\end{subequations}
where $\mathbf{P}$ is the closed-shell (spatial orbital) density matrix and 
$\mathbf{K}$ is the (symmetric, positive semi-definite) closed-shell anomalous 
density matrix.  We emphasize here that only the $\alpha\beta$ and 
$\beta\alpha$ blocks of $\bm{\kappa}$ are non-zero, so that we consider only 
singlet pairing.\cite{SS2002}  We should also mention that the notation here 
differs slightly from that used in Refs. \onlinecite{CPMFT1,CPMFT2,CPMFT3}, 
but does so in an attempt to make this manuscript self-contained and as clear 
as possible.

Idempotence of the quasiparticle density matrix $\mathbf{R}$ yields two 
conditions on the electronic density matrix $\mathbf{P}$ and the anomalous 
density matrix $\mathbf{K}$:
\begin{subequations}
\begin{align}
\mathbf{P} \, \mathbf{K} - \mathbf{K} \, \mathbf{P} &= \bm{0}
\label{Eqn:PKCommute}
\\
\mathbf{P} - \mathbf{P}^2 &= \mathbf{K}^2.
\label{Eqn:KapIdem}
\end{align}
\end{subequations}
Physically, $\mathbf{K}^2$ is the ``odd-electron distribution'' of 
Yamaguchi,\cite{Yamaguchi}  the ``density of effectively unpaired electrons'' 
of Staroverov and  Davidson,\cite{Star-David} and is related to Mayer's ``free 
valence index''\cite{Mayer} once it is written in terms of the total density 
matrix 
$\bm{\gamma}^{\alpha\alpha} + \bm{\gamma}^{\beta\beta} = 2 \mathbf{P}$.  
Essentially, $\mathbf{K}^2$ gauges the singlet diradical character of the 
system (or, for larger active spaces, the polyradical character) and is a 
local measure of electron entanglement.

The HFB energy is given as the expectation value of the Hamiltonian with 
respect to the HFB wave function
\begin{subequations}
\begin{align}
E_\mathrm{HFB} &= \langle \Phi_\mathrm{HFB} | H | \Phi_\mathrm{HFB} \rangle
\\
               &= 2 h_{ij} P_{ij} + (2 \langle ij | kl \rangle - \langle ij | lk \rangle)  P_{ik} P_{jl} 
\label{EHFB}
\\
               &\qquad+\langle ij | kl \rangle  K_{ij} K_{kl}
\nonumber
\end{align}
\end{subequations}
where summation over repeated indices here and throughout the manuscript is 
implied; $h_{ij}$ are matrix elements of the one-electron part of the 
Hamiltonian and  $\langle ij | kl \rangle$ are two-electron integrals in Dirac 
notation.

In order to determine the occupation numbers and natural orbitals, HFB 
variationally minimizes $E_\mathrm{HFB}$ subject to the constraint that the 
density matrix $\mathbf{P}$ contains the correct number of particles:
\begin{equation}
\mathrm{Tr}(\mathbf{P}) = N.
\end{equation}
This condition is enforced by a chemical potential $\mu$ introduced as a
Lagrange multiplier. 
The HFB formulation leads to equations similar to Hartree-Fock, 
which in the particular case of closed-shell systems are
\begin{equation}
\mathbf{R}_\mathrm{cs} \, \bm{\mathcal{H}}_\mathrm{HFB} - \bm{\mathcal{H}}_\mathrm{HFB} \, \mathbf{R}_\mathrm{cs} = \bm{0},
\end{equation}
where $\mathbf{R}_\mathrm{cs}$ is the closed shell quasiparticle density 
matrix 
\begin{equation}
\mathbf{R}_\mathrm{cs} = 
\begin{pmatrix} 
\mathbf{P} & \mathbf{K} \\ \mathbf{K} & \bm{1}-\mathbf{P}
\end{pmatrix}
\end{equation}
and $\bm{\mathcal{H}}_\mathrm{HFB}$ is the double-Hamiltonian (DH) given by 
\begin{equation}
\bm{\mathcal{H}}_\mathrm{HFB} = 
\begin{pmatrix} 
\mathbf{F}^\mathrm{cs} + \mu N & \bm{\Delta} \\ \bm{\Delta} & -\mathbf{F}^\mathrm{cs} - \mu N 
\end{pmatrix}.
\end{equation}
Here $\mathbf{F}^\mathrm{cs}$ is the standard closed-shell Fock matrix and 
$\bm{\Delta}$ is known as the pairing Hamiltonian.  These are given by
\begin{subequations}
\begin{align}
F^\mathrm{cs}_{ij} &= h_{ij} + (2 \langle ik|jl \rangle - \langle ik|lj \rangle) P_{kl},
\\
\Delta_{ij} &= \langle ij | kl \rangle K_{kl}.\label{Eqn:Delta}
\end{align}
\end{subequations}
The double-Hamiltonian $\bm{\mathcal{H}}_\mathrm{HFB}$ is just the 
mean-field of the physical Hamiltonian with respect to the quasiparticle 
determinant.

Because the pairing energy of HFB (the term proportional to $\mathbf{K}^2$ in 
Eqn. (\ref{EHFB})) is positive when the electron-electron interaction is 
repulsive, the variationally optimal solution is always $\mathbf{K} = \bm{0}$ 
and therefore $\bm{\Delta} = \bm{0}$.  In other words, HFB just returns the 
regular Hartree-Fock solution for Coulombic repulsive systems.  In order to 
have HFB solutions with energies lower than Hartree-Fock, one needs a net 
attractive two-body interaction, as in the Bardeen-Cooper-Schriefer picture of 
superconductivity (where it is provided by electron-phonon coupling) or in 
nuclear forces.  In order to take advantage of the pairing picture for the 
conventional repulsive electron-electron interaction, and with the aim of 
describing strong correlations, CPMFT simply reverses the sign of the pairing 
energy.  We thus have
\begin{equation}
\begin{split}
E_\mathrm{CPMFT} &= 2 h_{ij} P_{ij} + (2 \langle ij | kl \rangle - \langle ij | lk \rangle)  P_{ik} P_{jl}
\\
                 & \qquad- \langle ij | kl \rangle  K_{ij} K_{kl}
\label{Eqn:ECPMFT}
\end{split}
\end{equation}
The last term plays the role of a correlation energy -- a correction to the 
closed shell RHF-like energy expression -- and will be referred to as such 
throughout this manuscript, but it is certainly not our previous definition of 
static correlation,\cite{CPMFT2} $E_\mathrm{CPMFT} - E_\mathrm{RHF}$, since 
$\mathbf{P}$ is not $\mathbf{P}_\mathrm{RHF}$.

In addition to changing the sign of the pairing energy, in CPMFT we also 
restrict non-integer occupations to an active space, so that pairing only 
occurs between quasidegenerate orbitals.  Changing the sign of the pairing 
term changes the sign of $\bm{\Delta}$ so that the 
double-Hamiltonian is
\begin{equation}
\bm{\mathcal{H}}_\mathrm{CPMFT} = 
\begin{pmatrix} 
\mathbf{F}^\mathrm{cs} + \mu N & -\bm{\Delta} \\ -\bm{\Delta} & -\mathbf{F}^\mathrm{cs} - \mu N 
\end{pmatrix}.
\label{Eqn:DHCPMFT}
\end{equation}
Otherwise, DH-CPMFT follows the same procedure as in HFB.  However, changing 
the sign of the pairing energy and the pairing matrix severs the connection 
between the HFB wave function $|\Phi_\mathrm{HFB}\rangle$ and the CPMFT 
energy.  Note that what we have called simply CPMFT in Refs. 
\onlinecite{CPMFT1,CPMFT2,CPMFT3} is here referred to as DH-CPMFT, whereas 
``CPMFT'' here refers to the new formulation to be introduced below.

We can, indeed, view the CPMFT energy as the expectation value of a model 
Hamiltonian with respect to a particle-number violating determinant:
\begin{subequations}
\begin{align}
H_0 | \Phi \rangle &= E_\mathrm{CPMFT} |\Phi\rangle,
\\
H_0  &= (F^\mathrm{cs}_{ij} + h_{ij}) a_i^\dagger a_j 
\\
     &- \frac{1}{2}\Delta_{ij} a_i^\dagger a_j^\dagger - \frac{1}{2} \Delta^\star_{ij} a_i a_j.
\nonumber
\end{align}
\end{subequations}
This quadratic model Hamiltonian, however, is \textit{not} the mean-field of 
the physical Hamiltonian with respect to a quasiparticle determinant.  As 
previously noted,\cite{CPMFT2} we can interpret the CPMFT energy as a hybrid 
of Hartree-Fock and HFB where Hartree-Fock uses $2/r_{12}$ as the 
electron-electron repulsion operator and HFB uses $-1/r_{12}$.

Nevertheless, we have a fruitful alternative viewpoint, which is to envision 
the CPMFT energy expression of Eqn. (\ref{Eqn:ECPMFT}) as defining a model 
two-particle density matrix $\bm{\Gamma}$ such that the energy in the 
spin-orbital basis is 
\begin{equation}
E_\mathrm{CPMFT} = \mathrm{Tr}(\mathbf{h}\,\bm{\gamma}) + \mathrm{Tr}(\mathbf{v} \, \bm{\Gamma}_\mathrm{CPMFT})
\end{equation}
where $\mathbf{v}$ is the two-particle part of the Hamiltonian, and 
$\mathbf{h}$ is the one-particle part.  In terms of spin-orbitals, we have
\begin{equation}
(\bm{\Gamma}_\mathrm{CPMFT})_{ij}^{kl} = \frac{1}{2} \gamma_i^k \gamma_j^l -  \frac{1}{2} \gamma_i^l \gamma_j^k - \frac{1}{2} \kappa_{ij} \kappa^{kl}
\end{equation}
with lower (upper) indices corresponding to bra (ket) indices.  The first two 
terms in this model two-particle density matrix correspond to Hartree-Fock 
whereas the last term introduces static correlation via $\mathbf{K}$, which is 
a measure of non-idempotency for $\mathbf{P}$.  This last term is an important 
quantity in the cumulant decomposition of density matrices,\cite{Kutzelnigg} 
but in our work appears naturally from the idempotency of the quasiparticle 
density matrix. If we use this model two-particle density matrix to 
\textit{define} expectation values of two-particle operators, then as shown in 
the Appendix, we find the important result that CPMFT has no particle 
number fluctuations.  In making this choice, we are 
inevitably working with a density matrix functional and are effectively doing 
some form of a statistical ensemble theory.  Table \ref{Tab:Properties} 
collects results about the UHF two-particle density matrix, the CPMFT model 
two-particle density matrix, and the analogously defined HFB model 
two-particle density matrix.  We derive these results in the Appendix.

\begin{table}
\caption{Summary of properties in UHF, HFB, and CPMFT for closed-shell 
systems.  We show the correlation energy (\textit{i.e} the difference between 
the energy from the method and the closed-shell piece), the effective 
polarization, and the particle number fluctuations.
\label{Tab:Properties}}
\begin{tabular}{lccc}
\hline\hline
Method   &  $E_c$\footnote{~$v^{ij}_{kl} = \langle ij | kl \rangle$ is a 
two-electron integral in Dirac notation}
         &  Polarization
         &  $\sigma_N^2$		\\
\hline
UHF      & $-v_{ij}^{kl} \, M^i_l \, M^j_k$
         & $\mathbf{M} = (\mathbf{A} - \mathbf{B})/2$
         & 0\\
HFB      & $v_{ij}^{kl} \, K^{ij} \, K_{kl}$
         & $\mathbf{K} = |\mathbf{A} - \mathbf{B}|/2$
         & $4 \, \mathrm{Tr} (\mathbf{K}^2)$\\
CPMFT    & $-v_{ij}^{kl} \, K^{ij} \, K_{kl}$
         & $\mathbf{K} = |\mathbf{A} - \mathbf{B}|/2$
         & 0\\
\hline\hline
\end{tabular}
\end{table}

Note that our model two-particle density matrix has appeared before in the 
literature as the corrected Hartree-Fock (CHF) functional.\cite{CHF} 
Our model is, however, solved by diagonalization.\cite{SS2002}  More 
importantly, CPMFT restricts the non-integer occupation numbers to an active 
space only, and in the present work further enforces the corresponding pairs 
constraint, which we will now introduce.

\section{CPMFT and UHF 
\label{Sec:UHF}}
Consider the UHF treatment of a system where the number of spin-up and 
spin-down electrons is the same.  The spin-up and spin-down density matrices 
$\bm{\gamma}^{\alpha\alpha}$ and $\bm{\gamma}^{\beta\beta}$ 
are both idempotent:
\begin{equation}
(\bm{\gamma}^{\alpha\alpha})^2 - \bm{\gamma}^{\alpha\alpha} = 
(\bm{\gamma}^{\beta\beta})^2   - \bm{\gamma}^{\beta\beta}   = \bm{0}.
\end{equation}
The charge density and spin magnetization (or polarization) density matrices 
are
\begin{subequations}
\begin{align}
\mathbf{P} &= \frac{1}{2} \big(\bm{\gamma}^{\alpha\alpha} + \bm{\gamma}^{\beta\beta}\big)
\\
\mathbf{M} &= \frac{1}{2} \big(\bm{\gamma}^{\alpha\alpha} - \bm{\gamma}^{\beta\beta}\big).
\end{align}
\label{Eqn:DefPM}
\end{subequations}
Traditionally, the UHF energy\cite{Pople}  is expressed in terms of the
$\bm{\gamma}^{\alpha\alpha}$ and $\bm{\gamma}^{\beta\beta}$ density matrices:
\begin{align}
E_\mathrm{UHF} &= h_{ij} (\gamma^{\alpha\alpha}_{ij} + \gamma^{\beta\beta}_{ij})  
\\
               &+ \frac{1}{2} \langle ij | kl \rangle (\gamma^{\alpha\alpha}_{ik} + \gamma^{\beta\beta}_{ik}) (\gamma^{\alpha\alpha}_{jl} + \gamma^{\beta\beta}_{jl}) 
\nonumber
\\
&
- \frac{1}{2} \langle ij | kl \rangle (\gamma^{\alpha\alpha}_{il} \gamma^{\alpha\alpha}_{jk} + \gamma^{\beta\beta}_{il} \gamma^{\beta\beta}_{jk})
\nonumber
\end{align}
where we have put $\bm{\gamma}^{\alpha\alpha}$ and $\bm{\gamma}^{\beta\beta}$ 
in the same basis (say, the atomic orbital basis).  Although it is almost 
never  presented in this way, we can also write the UHF energy as a functional 
of $\mathbf{P}$ and $\mathbf{M}$, which yields
\begin{subequations}
\begin{align}
E_\mathrm{UHF}[\mathbf{P},\mathbf{M}] &= E_\mathrm{cs}[\mathbf{P}] + E_c[\mathbf{M}],
\\
E_\mathrm{cs}[\mathbf{P}]  &= 2 h_{ij} P_{ij} 
\\
                           &+ (2 \langle ij | kl \rangle - \langle ij | lk \rangle )P_{ik} P_{jl}
\nonumber
\\
E_c[\mathbf{M}]            &= -\langle ij | kl \rangle M_{il} M_{jk}.
\end{align}
\label{Eqn:EUHF}
\end{subequations}
Here, $E_\mathrm{cs}$ indicates the usual RHF energy expression given in terms 
of the charge density matrix $\mathbf{P}$, while $E_c$ carries the correlation 
energy in terms of the spin magnetization density matrix $\mathbf{M}$.  An 
utterly unexpected result is that the closed-shell CPMFT energy expression of 
Eqn. (\ref{Eqn:ECPMFT}) is identical to the UHF energy expression of Eqn. 
(\ref{Eqn:EUHF}), except that the spin density matrix $\mathbf{M}$ is replaced 
by the anomalous density matrix $\mathbf{K}$.\cite{FOOTNOTE}  In cases in 
which UHF predicts static correlation by breaking symmetry (\textit{i.e} 
non-zero spin density),\cite{Fukutome} $\mathbf{P}$ is not idempotent.  
Instead, it satisfies
\begin{subequations}
\begin{align}
\mathbf{P} - \mathbf{P}^2 
   &= \frac{1}{2} (\bm{\gamma}^{\alpha\alpha} + \bm{\gamma}^{\beta\beta}) - \frac{1}{4} (\bm{\gamma}^{\alpha\alpha} + \bm{\gamma}^{\beta\beta})^2
\\
   &=  \frac{1}{4} (\bm{\gamma}^{\alpha\alpha} - \bm{\gamma}^{\beta\beta})^2
\\
   &= \mathbf{M}^2.
\end{align}
\label{Eqn:Idempotency}
\end{subequations}
This is one consequence of the idempotence of $\bm{\gamma}^{\alpha\alpha}$ and 
$\bm{\gamma}^{\beta\beta}$.  The second is 
\begin{equation}
\mathbf{P} \, \mathbf{M} + \mathbf{M} \, \mathbf{P} = \mathbf{M}.
\label{Eqn:PMCommute}
\end{equation}
Note that the condition of Eqn. (\ref{Eqn:Idempotency}) is the same as the 
CPMFT condition of Eqn. (\ref{Eqn:KapIdem}), again with $\mathbf{M}$ taking 
the role of $\mathbf{K}$.  Both the magnetization density matrix $\mathbf{M}$ 
and the anomalous density matrix $\mathbf{K}$ are Hermitian.  

While CPMFT and UHF thus use the same energy expression (one with $\mathbf{K}$ 
and the other with $\mathbf{M}$), $\mathbf{K}$ and $\mathbf{M}$ are not the 
same even though with the same density matrix $\mathbf{P}$, we have 
$\mathbf{K}^2 = \mathbf{M}^2$.  There are also some other important 
differences.  Both UHF and CPMFT impose an additional condition on these two 
matrices, which in UHF is given in Eqn. (\ref{Eqn:PMCommute}) while in CPMFT 
is instead given in Eqn. (\ref{Eqn:PKCommute}).  Additionally, $\mathbf{K}$ is 
positive semi-definite while $\mathbf{M}$ is traceless (and thus has both 
positive and negative eigenvalues).  Finally, because in UHF we write 
$\mathbf{P}$ as the half-sum of two idempotent matrices, its eigenvalues occur in 
what is known as ``corresponding pairs'' $n_i$ and 
$1-n_i$,\cite{Amos,Harriman} a terminology that we here adopt.  

That UHF has the corresponding pairs property has little to do with UHF 
\textit{per se}.  It originates simply from the observation\cite{math} that 
the eigenvalues of a matrix that is the half-sum of two idempotent matrices are 0, 
1, $\tfrac{1}{2}$, or a corresponding pair ($n,1-n$). Similarly, the 
eigenvalues of a matrix written as the half-difference of two idempotent matrices 
are 0, $\pm \tfrac{1}{2}$, or a corresponding pair 
($-n,n$).\cite{FOOTNOTE2}  Thus, for example, $\mathbf{M}$ has eigenvalues 
adding to 0 in pairs while $\mathbf{P}$ has eigenvalues adding to 1 in pairs.  
Quite generally, any non-integer eigenvalues of the charge density matrix from 
a single determinant method will be either $\tfrac{1}{2}$ or occur in a 
corresponding pair.  Eigenvalues of $\tfrac{1}{2}$ could be part of a 
corresponding pair (for entangled electrons) or may occur singly for open 
shells.  We should be clear that while matrices written as the sum of two 
idempotent matrices exhibit the corresponding pairs property, the converse 
is not necessarily true; a matrix whose eigenvalues come in corresponding 
pairs may or may not be the sum of two idempotents.

Unlike UHF, the eigenvalues of $\mathbf{P}$ in DH-CPMFT do not occur in 
corresponding pairs (except when the active space consists of two spatial 
orbitals).  That said, the corresponding pairs property has some attractive 
features for CPMFT.  Most important is that it eliminates overcorrelation 
between orbital pairs in different symmetries. This is ubiquitous for example 
in N$_2$ where the variational principle drives occupancy into orbitals at low 
energies and one must introduce multiple chemical potentials to retain the 
correct total number of $\sigma$ and $\pi$ electrons. A corresponding pair 
constraint controls this unphysical ``spilling'' and has the inherent 
attractive feature of limiting strong correlations to be an affair between 
orbital pairs.

Previously, we had introduced the corresponding pairs feature within the 
DH-CPMFT framework using different chemical potentials (Lagrange 
multipliers) for different irreducible representations of the system.  
However, in the general case where no spatial symmetry is present, imposition 
of this constraint leads to one Lagrange multiplier per orbital pair and a 
rather complicated nonlinear optimization problem. A more satisfactory and 
much simpler approach, however, is to write the CPMFT density matrix as 
\begin{equation}
\mathbf{P} = \frac{1}{2} \left(\mathbf{A} + \mathbf{B}\right)
\label{Eqn:DefP}
\end{equation}
where $\mathbf{A}$ and $\mathbf{B}$ are auxiliary density matrices, 
individually idempotent and Hermitian
($\mathbf{A}^2 = \mathbf{A}= \mathbf{A}^\dagger$ and similarly for 
$\mathbf{B}$).  As with UHF, the decomposition above enforces the 
corresponding pairs condition automatically, and there is no need to enforce 
this condition via  Lagrange multipliers.  Eigenvalues of 0 or 1 in 
$\mathbf{P}$ correspond to virtual or core orbitals, respectively, while 
paired eigenvalues correspond to active orbitals.  Further, by choosing 
$\mathbf{A}$ and $\mathbf{B}$ to trace to half the number of electrons, we 
guarantee that $\mathbf{P}$ does likewise, and we thus have no need of any 
chemical potential.  By making this decomposition, in other words, we can 
avoid the Lagrange multipliers of the double-Hamiltonian approach entirely, 
and thus simplify the computation.  Note that once we have converged solutions 
for $\mathbf{A}$ and $\mathbf{B}$ (and thus $\mathbf{P}$ and $\mathbf{K}$), we 
could, if desired, extract the Lagrange multipliers of the DH-CPMFT approach.

The critical mathematical difference between CPMFT as formulated in this 
manner and UHF is that in UHF, we get $\mathbf{M}$ from the spin-up and 
spin-down density matrices, while in CPMFT, we get $\mathbf{K}$ from the total 
density matrix alone (since $\mathbf{K}$ satisfies the condition of Eqn. 
(\ref{Eqn:KapIdem}), commutes with $\mathbf{P}$, and is positive 
semi-definite).  In other words, CPMFT with corresponding pairs defines 
$\mathbf{P}$ from $\mathbf{A}$ and $\mathbf{B}$ as in Eqn. (\ref{Eqn:DefP}), 
but differs from UHF in constructing
\begin{equation}
\mathbf{K} = \sqrt{\mathbf{P} - \mathbf{P}^2} = \frac{1}{2} \sqrt{(\mathbf{A} - \mathbf{B})^2} = \frac{1}{2} | \mathbf{A} - \mathbf{B} |.
\label{Eqn:DefK}
\end{equation}
from auxiliary density matrices $\mathbf{A}$ and $\mathbf{B}$ while UHF 
builds $\mathbf{P}$ and $\mathbf{M}$ from $\bm{\gamma}^{\alpha\alpha}$ and 
$\bm{\gamma}^{\beta\beta}$, as shown in Eqn. (\ref{Eqn:DefPM}). Note in the 
last equation our definition of the absolute value of a matrix from the square 
root of the square. In practice, to calculate the absolute value of a matrix 
one needs to diagonalize it, flip the sign of the negative eigenvalues 
and transform back to the original basis. Both the square root and absolute 
value of a matrix are positive definite matrices and both have a convergent 
polynomial series expansion if the matrix is positive definite with 
eigenvalues between 0 and 1, as is the case here.

To make the comparison between CPMFT and UHF more concrete, consider the case 
where $\mathbf{A}$ and $\mathbf{B}$ are 2$\times$2 matrices and let 
$\mathbf{M} = \frac{1}{2} (\mathbf{A}-\mathbf{B})$.  Idempotency of 
$\mathbf{A}$ and $\mathbf{B}$ requires that in the natural orbital basis we 
have
\begin{subequations}
\begin{align}
\mathbf{A} &= 
   \begin{pmatrix} n & k \\ k & 1-n \end{pmatrix}
\\
\mathbf{B} &=
   \begin{pmatrix} n & -k \\ -k & 1-n \end{pmatrix}
\\
\mathbf{P} &= 
   \begin{pmatrix} n & 0 \\ 0 & 1-n \end{pmatrix}
\\
\mathbf{M} &= 
   \begin{pmatrix} 0 & k \\ k & 0 \end{pmatrix}
\\
\mathbf{K} &=
   \begin{pmatrix} k & 0 \\ 0 & k \end{pmatrix}
\\
k &= \sqrt{n(1-n)}.
\end{align}
\end{subequations}
When $\mathbf{A}$ and $\mathbf{B}$ are of larger dimension, then in the 
natural orbital basis they are block diagonal with  2$\times$2 blocks of the 
form given above.  This is essentially a consequence of Eqn. 
(\ref{Eqn:PMCommute}), which in the natural basis becomes
\begin{equation}
(n_i + n_j) M_{ij} = M_{ij} = \frac{1}{2} (A_{ij} - B_{ij}),
\end{equation}
the solutions to which are $M_{ij} = 0$ and $n_i + n_j = 1$.  Because we also 
have $A_{ij} + B_{ij} = 2 \, n_i \, \delta_{ij}$, we conclude that for 
$i \neq j$, we must either have $A_{ij} = B_{ij} = 0$ or $n_i + n_j = 1$ (in 
other words, the two eigenvalues form a corresponding pair).  When the 
occupation numbers are degenerate, the natural orbitals are not uniquely 
defined and we can thus choose them such that $\mathbf{A}$ and $\mathbf{B}$ 
still have this structure.  In the core (virtual) space, 
$\mathbf{A} = \mathbf{B} = \bm{1}$ ($\mathbf{A} = \mathbf{B} = \bm{0}$).

Before we continue to the working equations for CPMFT in this UHF-like 
framework, let us pause to make it explicit that CPMFT and UHF are different 
methods. While we have expressed the UHF energy as a density matrix 
functional, we could also write it as an expectation value
\begin{equation}
E_\mathrm{UHF} = \langle \Phi_\mathrm{UHF} | H | \Phi_\mathrm{UHF} \rangle
\end{equation}
with $|\Phi_\mathrm{UHF}\rangle$ constrained to be a single determinant.  
This is not true of the CPMFT energy expression, and in fact there seems to be 
no wave function associated with CPMFT.  This may seem somewhat surprising, in 
light of the intimate connection between CPMFT and HFB theory, in which there 
certainly \textit{is} a wave function, albeit one which violates particle 
number conservation.  As we have said, we lose the HFB wave function because 
we have by fiat changed the sign of the pairing energy.  Additionally, unlike 
UHF, the spin density is zero everywhere for closed shells, even in the presence 
of static correlation.

One might wonder whether CPMFT is equivalent to projected UHF (PUHF).  It is 
not.  If one projects the UHF determinant onto a spin eigenfunction, one finds 
that the charge density matrix of the UHF determinant and the spin-projected 
state have the same eigenfunctions.\cite{Harriman}  Spin projection, in other 
words, changes only the occupation numbers of the charge density matrix, but 
not the natural orbitals.  The fact that the UHF and CPMFT natural orbitals 
are different should lay to rest any concerns that CPMFT is just a projected 
UHF.

Another fundamental difference between CPMFT and UHF is the onset of the 
appearance of the solution with energy lower than RHF. As shown in our 
previous paper,\cite{CPMFT2} the CPMFT solution for a two-level model system 
appears when the RHF orbital energy gap reduces to
\begin{equation}
\varepsilon_2 - \varepsilon_1 <  \frac{1}{2} \langle 11|11 \rangle + \frac{1}{2} \langle 22|22 \rangle + \langle 11|22 \rangle,
\label{eq:CF1}
\end{equation}
whereas the UHF Coulson-Fischer instability point is determined by
\begin{equation}
\varepsilon_2 - \varepsilon_1 <  \langle 12|12 \rangle + \langle 11|22 \rangle .
\label{eq:CF2}
\end{equation}
Because all two-electron integrals in the equations above are positive, the 
CPMFT solution appears inevitably when the orbital gap closes and strong 
correlation is manifest, such as along a dissociation curve.

\subsection{Working Equations}
Let us now return to the solution of the CPMFT equations in this UHF-like 
framework.  For convenience, we repeat the energy expression here:
\begin{equation}
E_\mathrm{CPMFT} = E_\mathrm{cs} -  \langle ij  | kl \rangle K_{ij} K_{kl}.
\label{WorkingEqns}
\end{equation}
We simply minimize the energy with respect to (idempotent) $\mathbf{A}$ and 
$\mathbf{B}$ matrices.  The derivatives of $E_\mathrm{cs}$ in Eqn. 
(\ref{WorkingEqns}) with respect to $\mathbf{A}$ and $\mathbf{B}$ give the 
usual closed-shell Fock matrix obtained from $\mathbf{P}$.  That is
\begin{equation}
            \frac{\partial E_\mathrm{cs}}{\partial A_{ij}\hfill} = 
            \frac{\partial E_\mathrm{cs}}{\partial B_{ij}\hfill} = 
\frac{1}{2} \frac{\partial E_\mathrm{cs}}{\partial P_{ij}\hfill} = 
F_{ij}^\mathrm{cs}.
\end{equation}

The differences with UHF arise from differentiating the last term of the CPMFT 
energy, which we shall call $E_c^\mathrm{CPMFT}$.  Taking derivatives with 
respect to $\mathbf{A}$ leads to an effective potential 
$\tilde{\bm{\Delta}}$, given by
\begin{equation}
\tilde{\Delta}_{ij}
= \frac{\partial E_c^\mathrm{CPMFT}}{\partial A_{ij}\hfill} 
= \frac{\partial E_c^\mathrm{CPMFT}}{\partial K_{kl}\hfill} \frac{\partial K_{kl}}{\partial A_{ij}} 
= -2\Delta_{kl} \frac{\partial K_{kl}}{\partial A_{ij}}.
\end{equation}
This is essentially the same result that we get from differentiating 
$E_c^\mathrm{UHF}$ of Eqn. (\ref{Eqn:EUHF}):
\begin{equation}
\frac{\partial E_c^\mathrm{UHF}}{\partial \gamma^{\alpha\alpha}_{ij}} = \frac{\partial E_c^\mathrm{UHF}}{\partial M_{kl}} \frac{\partial M_{kl}}{\partial \gamma^{\alpha\alpha}_{ij}} = -2 \Delta_{kl}^\mathrm{UHF} \frac{\partial M_{kl}}{\partial \gamma^{\alpha\alpha}_{ij}}
\end{equation}
where
\begin{equation}
\Delta_{kl}^\mathrm{UHF} = \langle km | nl \rangle M_{mn} = \langle kl | mn \rangle M_{mn}
\end{equation}
looks just like $\bm{\Delta}$ except we replace $\mathbf{K}$ with 
$\mathbf{M}$.  In UHF, however, we simply have
\begin{equation}
\frac{\partial M_{kl}}{\partial \gamma^{\alpha\alpha}_{ij}} = \frac{1}{2} \delta_{ik} \delta_{jl}
\end{equation}
while in CPMFT the derivative of $\mathbf{K}$ with respect to $\mathbf{A}$ is 
obtained by differentiating both sides of 
$\mathbf{K}^2 = \tfrac{1}{4} (\mathbf{A}-\mathbf{B})^2$.  This gives
\begin{equation}
\frac{\partial K_{km}}{\partial A_{ij}} K_{ml} + K_{km} \frac{\partial K_{ml}}{\partial A_{ij}} = \frac{1}{2} \left(M_{jl} \delta_{ki} + M_{ki} \delta_{jl}\right).
\end{equation}
In the natural orbital basis where $\mathbf{K}$ is diagonal with eigenvalues 
$K_i$, we have
\begin{equation}
\frac{\partial K_{kl}}{\partial A_{ij}} = \frac{1}{2} \frac{M_{jl}\delta_{ki} + M_{ki} \delta_{jl}}{K_k + K_l}.
\end{equation}
Thus, in the natural orbital basis the effective potential 
$\tilde{\bm{\Delta}}$ is
\begin{equation}
\tilde{\Delta}_{ij}
= -\frac{\Delta_{il} M_{jl}}{K_i + K_l} 
  -\frac{\Delta_{kj} M_{ki}}{K_k + K_j}.
\label{Eqn:Dtilde}
\end{equation}
Since 
\begin{equation}
\frac{\partial \mathbf{K}}{\partial \mathbf{A}} = -\frac{\partial \mathbf{K}}{\partial \mathbf{B}},
\end{equation}
the equations we ultimately solve are
$[\mathbf{F}^\mathrm{A},\mathbf{A}] = \bm{0}$ and 
$[\mathbf{F}^\mathrm{B},\mathbf{B}] = \bm{0}$, where 
$\mathbf{F}^\mathrm{A}$ and $\mathbf{F}^\mathrm{B}$ are 
effective Fock matrices given by
\begin{subequations}
\begin{align}
\mathbf{F}^\mathrm{A} &= \mathbf{F}^\mathrm{cs} + \tilde{\bm{\Delta}},
\\
\mathbf{F}^\mathrm{B} &= \mathbf{F}^\mathrm{cs} - \tilde{\bm{\Delta}}.
\end{align}
\end{subequations}

At first glance, the right-hand-side of Eqn. \ref{Eqn:Dtilde} might appear to 
be divergent unless all $K_i$ are non-zero.  However, since forcing 
$\Delta_{ij}=0$ actually gives the condition $K_{ij}=0$, we simply set 
$\Delta_{ij}=0$ for the inactive-inactive (core and virtual) block where 
$\mathbf{K}$ must be zero (because the occupation numbers are 0 or 1).  
Therefore, in Eqn. \ref{Eqn:Dtilde}, such divergent terms due to inactive 
orbitals are simply removed from the sum.

\section{Results
\label{Sec:Results}}
We have implemented this version of CPMFT in the \texttt{Gaussian} suite of 
programs.\cite{GAUSSIAN}  Each calculation requires the specification of the 
number $N_\mathrm{Act}$ of active natural orbitals.  Due to the corresponding 
pairs constraint, the number of active electrons is always equal to 
$N_\mathrm{Act}$ -- in other words, we always work at half-filling.  In order 
to obtain an appropriate initial guess for $\mathbf{A}$ and $\mathbf{B}$, we 
mix the coefficients of the $N_\mathrm{Act}$ orbitals closest to the Fermi 
level, just as one would do to break spatial symmetries in UHF. The natural 
orbital pairs closest to the Fermi energy correspond to those whose 
occupations are closest to half and half.

In single bond systems where we normally choose the active space to be two 
electrons in two orbitals, the corresponding pair constraint is automatically 
satisfied, and no difference is observed between the results using the present 
approach and those using our previous double-Hamiltonian approach (that is, 
diagonalization of the double-Hamiltonian constructed from $\mathbf{F}$ and 
$\bm{\Delta}$).  However, in DH-CPMFT, one must adjust the chemical potential 
$\mu$ at every iteration of the SCF procedure to control the number of 
electrons in the active space.  Because we must adjust the chemical potential, 
we must diagonalize the double Hamiltonian of Eqn. (\ref{Eqn:DHCPMFT}) 
several times in each SCF cycle, until the resulting density matrix has the 
proper trace.  In contrast, the current approach requires no chemical 
potential, since we have 
$\mathrm{Tr}(\mathbf{P}) = 1/2 \, \mathrm{Tr}(\mathbf{A} + \mathbf{B})$.  
Because both $\mathbf{A}$ and $\mathbf{B}$ trace to the correct number of 
electrons, so too does $\mathbf{P}$.  This is a significant operational 
advantage of the present implementation.

\begin{table*}[t]
\caption{CPMFT energies of N$_2$ at $R$ = 2.0 \AA.  Also included are the 
number of diagonalization steps required, $N_\mathrm{diag}$, and the number 
of SCF cycles required for convergence.
\label{Tab:N2}}
\begin{tabular}{lcrr}
\hline\hline
Scheme      & Energy ($a.u.$) &  $N_\mathrm{diag}$ &  SCF cycles   \\
\hline
DH-CPMFT(6,6)\footnote{Single chemical potential}
            & -108.79901762   &  118               &       32      \\
DH-CPMFT(6,6)\footnote{Corresponding pairs enforced by multiple chemical 
potentials}
            & -108.79715442   &  121               &       34      \\
CPMFT(6)    & -108.79715442   &   12               &       12      \\
\hline\hline
\end{tabular}
\end{table*}

\begin{figure}[t]
\includegraphics[width=0.45\textwidth]{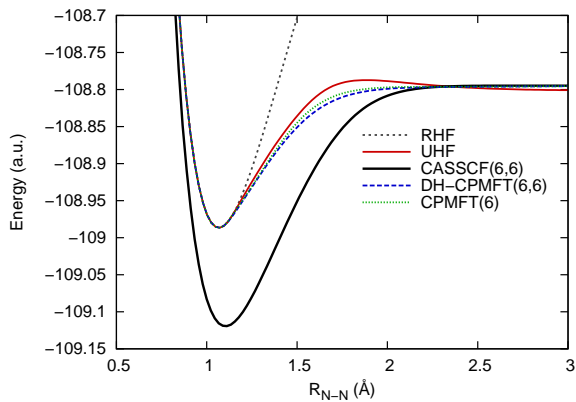}
\caption{Potential energy curves of N$_2$ calculated with the cc-pVTZ basis 
set.
\label{Fig:N2}}
\end{figure}

For systems with larger active spaces, the present approach differs from 
DH-CPMFT, although as mentioned above, we can impose the corresponding pairs 
constraint in DH-CPMFT in some special cases by including different chemical 
potentials for different irreducible representations.  We illustrate this with 
the case of N$_2$.  Table \ref{Tab:N2} shows the total energy of N$_2$ at 2.0 
{\AA}.  We use the cc-pVTZ basis set and choose six active orbitals and six 
active electrons.  The current scheme gives a slightly higher energy than does 
DH-CPMFT with only one chemical potential, as one would expect since we have 
imposed an additional constraint on the system.  Also as one would expect, it 
gives the same results as does DH-CPMFT with the corresponding pairs 
constraint enforced by additional Lagrange multipliers.  However, removing the 
chemical potentials results in considerable computational savings.  In Fig. 
\ref{Fig:N2}, we show the N$_2$ dissociation curves from CPMFT in the 
double-Hamiltonian approach and in the corresponding pairs framework.  In this 
case, the corresponding pairs constraint has only a minor effect on the energy.

\begin{figure}[t]
\includegraphics[width=0.45\textwidth]{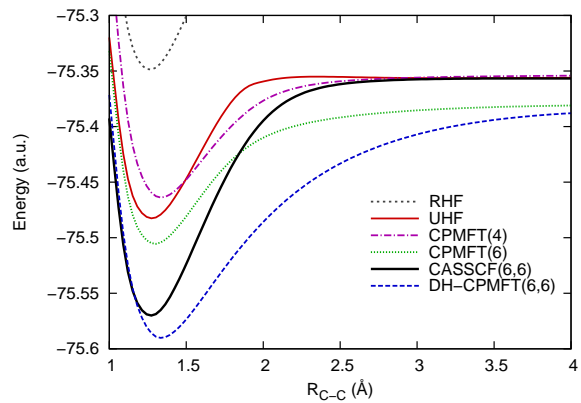}
\caption{Potential energy curves of C$_2$ calculated with the 6-31G basis set.
\label{Fig:C2}}
\end{figure}

We have also performed a CPMFT calculation of the C$_2$ molecule with the 
6-31G basis set.  Near equilibrium, C$_2$ has significant static correlation 
due to near-degeneracy between the RHF occupied $\sigma_{2s}^\star$ and 
unoccupied $\sigma_{2p_z}$ orbitals.  As the molecule is stretched, however, 
the $\pi_x$, $\pi_y$, $\pi_x^\star$, and $\pi_y^\star$ orbitals become 
degenerate, while the $\sigma_{2s}^\star$--$\sigma_{2p_z}$ interaction becomes 
weak.  We have therefore chosen our active space to be six electrons in 
six orbitals for this system.  In Fig. \ref{Fig:C2} we show the total energy 
of C$_2$ as a function of bond length.  The CASSCF energy includes all static 
correlation that results from these orbital interactions (plus some dynamical 
correlation).  Without the corresponding pairs constraint, DH-CPMFT strongly 
overcorrelates nearly everywhere.  Adding the corresponding pairs constraint 
significantly reduces this overcorrelation.  Near equilibrium, it gives 
results between UHF and CASSCF.  Unfortunately, it still overcorrelates as the 
molecule dissociates.  This is due to electron ``spilling'' between 
$\sigma_{2s}^\star$ and $\sigma_{2p_z}$ orbitals.  As $R \to \infty$, only the 
$\pi$ orbitals should be strongly correlated; including these $\sigma$ 
orbitals in the active space at large internuclear separation allows them to correlate 
and lower the energy unphysically.  If we remove two orbitals from the active 
space, we produce the curve marked CPMFT(4).  This goes to the correct 
dissociation limit, but undercorrelates at equilibrium where the active 
space should be larger.  The correct solution for this molecule involves 
introducing renormalized one-body potentials in CPMFT(6) that eliminate the spilling at 
dissociation,\cite{CPMFT2} an approach that we will discuss in a forthcoming 
article.  While going to the right dissociation limit is important, it is 
perhaps less critical than getting the correct behavior near equilibrium.  
Note that CPMFT(4) dissociates correctly to two ROHF carbon atoms, while UHF 
instead dissociates to two spin-contaminated UHF carbon atoms and CASSCF(6,6) 
has some dynamical correlation at dissociation.

\begin{figure}[t]
\includegraphics[width=0.45\textwidth]{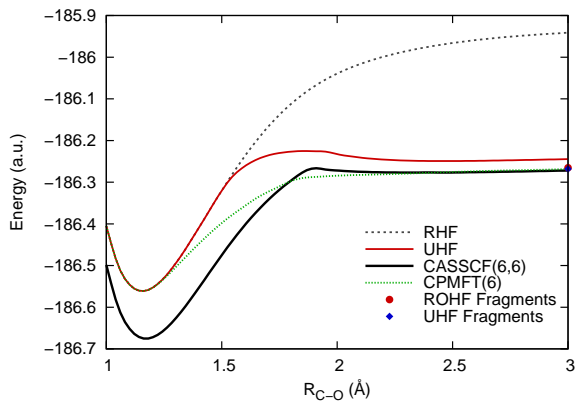}
\caption{Potential energy curves for the double dissociation of CO$_2$ 
calculated with the 3-21G basis set.
\label{Fig:CO2}}
\end{figure}

Finally, we stress the differences between UHF and CPMFT by analyzing the 
dissociation of the CO$_2$ molecule. The ground state of CO$_2$ near 
equilibrium is a closed-shell singlet with no expected static correlation. 
Indeed both UHF and CPMFT reduce to the RHF solution near R$_e$. However, when 
the molecule is symmetrically stretched and the two oxygen atoms are 
simultaneously separated from the carbon atom, the correct dissociation limit 
corresponds to all three atoms in their triplet ground state.  This situation 
cannot be handled by UHF. In CO$_2$ near R$_e$, there are six electrons 
associated with bond formation, three with spin-up and three with spin-down. 
At dissociation, UHF might assign two spin-up electrons to one oxygen atom and 
two spin-down electrons on the other, which puts both oxygen atoms in their 
triplet ground state.  However, with only one electron of each spin remaining, 
the best UHF can do is to assign a singlet state to the carbon atom, which is 
clearly incorrect and not the lowest energy state. In simple words, UHF runs 
out of broken symmetry degrees of freedom (has only two) to model the 
dissociation of CO$_2$ (Fig. \ref{Fig:CO2}) and misses the correct 
dissociation limit by $\sim$ 20 milliHartrees.  The bumps in the dissociation 
curves correspond to crossings of different solutions to the respective SCF 
equations and we have plotted the lowest energy state at each $R$.  Because 
spin states are treated in CPMFT through an ``ensemble'' representation, one 
that yields zero spin magnetization density everywhere, the CPMFT solution for 
this dissociation has two half spins up and two half spins down on each of the 
three atoms, leading to the correct energy corresponding to the sum of ROHF 
atomic energies.  Note that CPMFT(6) in Fig. \ref{Fig:CO2} contains a one-body 
potential arising from an asymptotic constraint as explained in our previous 
publication.\cite{CPMFT2}  We defer detailed discussion of the renormalization 
schemes used in CO$_2$ and applicable to C$_2$ within the current UHF-like 
context to a forthcoming publication. 

\section{Conclusions
\label{Sec:Conclusions}}
We have developed a novel scheme for performing CPMFT calculations with 
occupation numbers occurring in corresponding pairs.  In doing so, we 
eliminate all chemical potentials, and the effective Fock matrices 
$\mathbf{F}^\mathrm{A}$ and $\mathbf{F}^\mathrm{B}$ that are to be 
diagonalized are of half the dimension of the double Hamiltonian matrix in the 
previous DH-CPMFT scheme.  Thus, the computational effort in our present 
implementation is greatly reduced over the previous formulation of CPMFT.  The 
corresponding pairs constraint reduces the overcorrelation of C$_2$ near 
equilibrium, and has important consequences for the dissociation of 
heteronuclear systems.  While the corresponding pair constraint could also be 
imposed in the DH-CPMFT framework by addition of one Lagrange multiplier per 
electron pair, the current approach imposes this constraint in a simpler 
black-box manner.

We have shown that this version of CPMFT is closely related to UHF 
theory.  Unlike UHF, however, CPMFT incorporates static correlation by a different 
mechanism. The physical density matrix $\bm{\gamma}$ has 
identical spin-up and spin-down blocks, whereas the auxiliary $\mathbf{A}$ and 
$\mathbf{B}$ density matrices, in general, break symmetry. CPMFT can 
correctly dissociate polyatomic molecules into ROHF atoms or fragments, 
whereas UHF has problems with multiple entangled electrons at multiple
centers, as shown for CO$_2$ above. In the present formulation, CPMFT becomes 
a density matrix functional that can be solved by diagonalization of effective 
Fock matrices providing orbitals and orbital energies.  We wish to emphasize 
one more time that as we have demonstrated, a quasiparticle picture of strong 
correlations with the sign of the pairing interaction reversed yields an 
energy expression reminiscent of UHF.

Finally, we should note that in CPMFT different auxiliary $\mathbf{A}$ and 
$\mathbf{B}$ density matrices can lead to solutions with degenerate energies.  
The key quantities determining the energy in the model are $\mathbf{P}$ and 
$\mathbf{K}$ and there is a many-to-one mapping between $\mathbf{A}$ and 
$\mathbf{B}$ on the one hand and $\mathbf{P}$ and $\mathbf{K}$ on the other.  
At dissociation, for example, solutions where $\mathbf{A}$ and $\mathbf{B}$ 
orbitals are localized and delocalized (roughly corresponding to UHF and RHF 
orbitals) are degenerate. The existence of additional degenerate solutions in 
CPMFT (compared to UHF) can lead to convergence difficulties as the active 
space becomes large. Efficient ways of dealing with the additional degrees of 
freedom provided by the auxiliary $\mathbf{A}$ and $\mathbf{B}$ matrices are 
currently under investigation.

\section{Acknowledgments}
This work was supported by NSF (Grant No. CHE-0807194), LANL Subcontract 
81277-001-10, and the Welch Foundation (Grant No. C-0036). A.S. acknowledges
support from ANR (Grant No. 07-BLAN-0272). We thank Carlos Jim\'enez-Hoyos and 
Jason Ellis for useful discussions and Dan Sorensen for pointing out Ref. 
\onlinecite{math}.

\appendix
\section{Properties of the CPMFT Model Two-Particle Density Matrix}
The CPMFT model two-particle density matrix is
\begin{equation}
(\bm{\Gamma}_\mathrm{CPMFT})_{ij}^{kl} = \frac{1}{2} \gamma_i^k \gamma_j^l - \frac{1}{2} \gamma_i^l \gamma_j^k - \frac{1}{2} \kappa_{ij} \kappa^{kl}.
\label{Def2MatCPMFT}
\end{equation}
where $i$, $j$, $k$, and $l$ are spin-orbitals and $\bm{\gamma}$ and 
$\bm{\kappa}$ are the density matrix and anomalous density matrix in the 
spin-orbital basis (\textit{i.e}, they are of dimension $2N \times 2N$, where 
$N$ is the size of the atomic orbital basis).  In general, $\bm{\gamma}$ is 
Hermitian and $\bm{\kappa}$ is antisymmetric.  When everything is real (which 
we take for simplicity; this does not affect our conclusions), the idempotent 
HFB quasiparticle density matrix is
\begin{equation}
\mathbf{R} = 
\begin{pmatrix} \bm{\gamma} & \bm{\kappa} \\ 
               -\bm{\kappa} & \bm{1} - \bm{\gamma} \end{pmatrix}.
\end{equation}
Idempotency tells us that
\begin{subequations}
\begin{align}
\bm{\gamma} \, \bm{\kappa} - \bm{\kappa} \, \bm{\gamma} &= \bm{0},
\\
\bm{\gamma}^2 - \bm{\kappa}^2 &= \bm{\gamma}.
\end{align}
\end{subequations}
We recall that for closed shells,\cite{SS2002}
\begin{subequations}
\begin{align}
\bm{\gamma} &= 
 \begin{pmatrix} \mathbf{P} & \bm{0} \\ \bm{0} & \mathbf{P} \end{pmatrix},
\\
\bm{\kappa} &= 
 \begin{pmatrix} \bm{0} & \mathbf{K} \\ -\mathbf{K} & \bm{0} \end{pmatrix},
\\
\bm{0} &= \mathbf{P} \, \mathbf{K} - \mathbf{K} \, \mathbf{P},
\\
\mathbf{P} &= \mathbf{P}^2 + \mathbf{K}^2.
\end{align}
\label{ClosedShellStuff}
\end{subequations}

We can define an analogous model two-particle density matrix for HFB, for 
which all the conditions on $\bm{\kappa}$, $\bm{\gamma}$, $\mathbf{K}$, and 
$\mathbf{P}$ are the same, but where
\begin{equation}
(\bm{\Gamma}_\mathrm{HFB})_{ij}^{kl} = \frac{1}{2} \gamma_i^k \gamma_j^l - \frac{1}{2} \gamma_i^l \gamma_j^k + \frac{1}{2} \kappa_{ij} \kappa^{kl}.
\label{Def2MatHFB}
\end{equation}

Finally, the UHF two-particle density matrix is
\begin{equation}
(\bm{\Gamma}_\mathrm{UHF})_{ij}^{kl} = \frac{1}{2} \gamma_i^k \gamma_j^l - \frac{1}{2} \gamma_i^l \gamma_j^k
\label{Def2MatUHF}
\end{equation}
where $\bm{\gamma}$ is idempotent.  We have
\begin{subequations}
\begin{align}
\bm{\gamma} &= 
\begin{pmatrix} \bm{\gamma}^{\alpha\alpha} & \bm{0}  \\
                \bm{0} & \bm{\gamma}^{\beta\beta}  \end{pmatrix}
= 
\begin{pmatrix} \mathbf{P} + \mathbf{M} & \bm{0} \\
                \bm{0} & \mathbf{P} - \mathbf{M} \end{pmatrix},
\\
\mathbf{P} &= \mathbf{P}^2 + \mathbf{M}^2,
\\
\mathbf{M} &= \mathbf{P} \, \mathbf{M} + \mathbf{M} \, \mathbf{P}.
\end{align}
\label{ClosedShellUHF}
\end{subequations}

\subsection{Partial Trace of the Two-Particle Density Matrix}
An important condition on the two-particle density matrix is that it traces to 
the one-particle density matrix.  That is, we must have
\begin{equation}
\Gamma_{ij}^{il} = \frac{N-1}{2} \gamma_j^l.
\label{PartialTrace}
\end{equation}
We remind the reader that repeated indices are to be summed.

The partial trace condition is satisfied by the UHF two-matrix and the CPMFT 
model two-matrix, but not by the HFB model two-matrix:
\begin{subequations}
\begin{align}
\Gamma_{ij}^{il} 
  &= \frac{1}{2} \left(\gamma_i^i \gamma_j^l - \gamma_i^l \gamma_j^i \mp \kappa_{ij} \kappa^{il}\right)
\\
  &= \frac{1}{2} \left[N \gamma_j^l - (\bm{\gamma}^2)_j^l \pm (\bm{\kappa}^2)_j^l \right]
\\
  &= \frac{1}{2} \left[N \gamma_j^l - (\bm{\gamma} + \bm{\kappa}^2)_j^l \pm (\bm{\kappa}^2)_j^l \right]
\\
  &= \frac{N-1}{2} \gamma_j^l - \frac{1}{2}\left[(\bm{\kappa}^2)_j^l \mp (\bm{\kappa}^2)_j^l \right].
\end{align}
\end{subequations}
Here, the top (bottom) sign in $\pm$ and $\mp$ corresponds to CPMFT (HFB), and 
we have used antisymmetry of $\bm{\kappa}$.  Explicitly, we have
\begin{subequations}
\begin{align}
(\bm{\Gamma}_\mathrm{CPMFT})_{ij}^{il} &= \frac{N-1}{2} \gamma_j^l.
\\
(\bm{\Gamma}_\mathrm{HFB})_{ij}^{il}   &= \frac{N-1}{2} \gamma_j^l - (\bm{\kappa}^2)_j^l.
\end{align}
\end{subequations}
Note that by $N$ we mean the trace of the one-particle density matrix 
$\bm{\gamma}$, which should be the number of particles in the system.

\subsection{Particle Number Fluctuations}
In order to work out particle number fluctuations, we need the expectation 
values of $\hat{N}$ and $\hat{N}^2$, with $\hat{N}$ the number operator, 
given as
\begin{equation}
\hat{N} = \delta_{pq} a_p^\dagger a_q.
\end{equation}
We have already noted that the expectation value of $\hat{N}$ is just 
$\mathrm{Tr}(\bm{\gamma})$.  The expectation value of $\hat{N}^2$ requires the 
two-particle density matrix:
\begin{subequations}
\begin{align}
\langle \hat{N}^2 \rangle
  &= \delta_{pq} \, \delta_{rs} \, \langle a_p^\dagger a_q a_r^\dagger a_s \rangle
\\
  &= \delta_{pq} \, \delta_{rs} \left( -\langle a_p^\dagger a_r^\dagger a_q a_s \rangle + \delta_{qr} \langle a_p^\dagger a_s \rangle \right)
\\
  &= \delta_{pq} \, \delta_{rs} \left(2 \, \Gamma_{pr}^{qs} + \delta_{qr} \gamma_p^s\right)
\\
  &= 2 \, \Gamma_{pr}^{pr} + \gamma_p^p.
\end{align}
\end{subequations}

If the two-particle density matrix obeys the partial trace condition, the 
particle number fluctuations are automatically zero.  This is thus true of UHF 
and of CPMFT.  However, HFB has particle number fluctuations:
\begin{equation}
\begin{split}
\langle \hat{N}^2 \rangle_\mathrm{HFB} &= (N-1) \gamma_j^j - 2 (\bm{\kappa}^2)_j^j + \gamma_j^j 
\\
&= N^2 - 2 \, \mathrm{Tr}(\bm{\kappa})^2
\end{split}
\end{equation}
implying that
\begin{equation}
\sigma_N^2 = \langle \hat{N}^2 \rangle - \langle \hat{N} \rangle^2 = -2 \, \mathrm{Tr}(\bm{\kappa}^2).
\end{equation}
Note that this is positive, as it should be, since 
$-\bm{\kappa}^2~=\bm{\gamma}~-~\bm{\gamma}^2$ and occupation numbers are 
between 0 and 1, inclusive.  In the closed-shell case, we have 
$\sigma_N^2 = 4 \, \mathrm{Tr}(\mathbf{K}^2)$.

\subsection{Spin Contamination}
Evaluating spin contamination is more complicated than evaluating particle 
number fluctuations, not least because we need an expression for 
$\langle \hat{S}^2 \rangle$ for a general two-particle density matrix 
$\bm{\Gamma}$.  We begin by noting that
\begin{subequations}
\begin{align}
\hat{S}^2 &= \hat{S}_x^2 + \hat{S}_y^2 + \hat{S}_z^2
\\
          &= \hat{S}_z + \hat{S}_z^2 + \hat{S}_{-}\hat{S}_{+},
\end{align}
\end{subequations}
where $\hat{S}_\pm$ is the spin raising/lowering operator.  We are interested 
here in the closed-shell case (\textit{i.e.} $N_\alpha = N_\beta$ with a 
block diagonal $\bm{\gamma}$).

In the closed-shell case, the contribution to $\langle \hat{S}^2 \rangle$ from 
the first term is zero.  We must evaluate the contribution from the next piece 
using our model two-particle density matrix.  We have
\begin{equation}
\hat{S}_z^2 = \underbrace{\sum_i \hat{s}_z(i)^2}_{\hat{X}_z} + \underbrace{\sum_{i \neq j} \hat{s}_z(i) \hat{s}_z(j)}_{\hat{Y}_z}.
\end{equation}
The first (second) term is a one-particle (two-particle) operator.  Note that we could also write
\begin{equation}
\hat{Y}_z = 2 \sum_{i > j} \hat{s}_z(i) \hat{s}_z(j)
\end{equation}
which explains the factor of 2 that might otherwise appear to be missing below.

Evaluating the contribution to $\langle \hat{S}_z^2 \rangle$ from $\hat{X}_z$ 
is straightforward, and we get just
\begin{equation}
\langle \hat{X}_z \rangle = \frac{1}{4} (N_\alpha + N_\beta) = \frac{1}{2} \mathrm{Tr}(\mathbf{P}).
\end{equation}
The nonzero matrix elements of $\hat{Y}_z$ are
\begin{subequations}
\begin{align}
(Y_z)^{i_\alpha j_\alpha}_{k_\alpha l_\alpha} &= \langle i_\alpha j_\alpha | \hat{Y}_z | k_\alpha l_\alpha \rangle = \frac{1}{2} \delta^i_k \delta^j_l,
\\
(Y_z)^{i_\alpha j_\beta}_{k_\alpha l_\beta} &= \langle i_\alpha j_\beta | \hat{Y}_z | k_\alpha l_\beta \rangle = -\frac{1}{2} \delta^i_k \delta^j_l,
\\
(Y_z)^{i_\beta j_\alpha}_{k_\beta l_\alpha} &= \langle i_\beta j_\alpha | \hat{Y}_z | k_\beta l_\alpha \rangle = -\frac{1}{2} \delta^i_k \delta^j_l,
\\
(Y_z)^{i_\beta j_\beta}_{k_\beta l_\beta} &= \langle i_\beta j_\beta | \hat{Y}_z | k_\beta l_\beta \rangle = \frac{1}{2} \delta^i_k \delta^j_l.
\end{align}
\end{subequations}
Here, we are working in an orthornomal basis set.

The relevant components of the CPMFT and HFB two-particle density matrices are
\begin{subequations}
\begin{align}
\Gamma_{i_\alpha j_\alpha}^{k_\alpha l_\alpha} &= \frac{1}{2} \left(\gamma_{i_\alpha}^{k_\alpha} \gamma_{j_\alpha}^{l_\alpha} - \gamma_{i_\alpha}^{l_\alpha} \gamma_{j_\alpha}^{k_\alpha}\right),
\\
\Gamma_{i_\alpha j_\beta}^{k_\alpha l_\beta} &= \frac{1}{2}  \left(\gamma_{i_\alpha}^{k_\alpha} \gamma_{j_\beta}^{l_\beta} \mp \kappa_{i_\alpha j_\beta} \kappa^{k_\alpha l_\beta}\right),
\\
\Gamma_{i_\beta j_\alpha}^{k_\beta l_\alpha} &= \frac{1}{2}  \left(\gamma_{i_\beta}^{k_\beta} \gamma_{j_\alpha}^{l_\alpha} \mp \kappa_{i_\beta j_\alpha} \kappa^{k_\beta l_\alpha}\right),
\\
\Gamma_{i_\beta j_\beta}^{k_\beta l_\beta} &= \frac{1}{2}  \left(\gamma_{i_\beta}^{k_\beta} \gamma_{j_\beta}^{l_\beta} - \gamma_{i_\beta}^{l_\beta} \gamma_{j_\beta}^{k_\beta}\right),
\end{align}
\end{subequations}
where the top (bottom) sign corresponds to CPMFT (HFB).

Contracting the density matrices with the matrix elements, we get
\begin{equation}
\langle \hat{Y}_z \rangle = \left(\frac{N_\alpha - N_\beta}{2}\right)^2 - \frac{1}{4} \mathrm{Tr}(\bm{\gamma}_{\alpha\alpha}^2 + \bm{\gamma}_{\beta\beta}^2 \mp \bm{\kappa}_{\alpha\beta}^2 \mp \bm{\kappa}_{\beta\alpha}^2) \end{equation}
where we have used antisymmetry of $\bm{\kappa}$.  Working in our closed-shell case, this reduces to
\begin{equation}
\langle \hat{Y}_z \rangle = - \frac{1}{2} \mathrm{Tr}(\mathbf{P}^2 \mp \mathbf{K}^2).
\end{equation}

In total, then, we find that $\langle \hat{S}_z^2 \rangle$ in CPMFT and HFB is given by
\begin{subequations}
\begin{align}
\langle \hat{S}_z^2 \rangle &= \frac{1}{2} \mathrm{Tr}(\mathbf{P} - \mathbf{P}^2 \pm \mathbf{K}^2)
\\
                            &= \frac{1}{2} \mathrm{Tr}(\mathbf{K}^2 \pm \mathbf{K}^2).
\end{align}
\end{subequations}
Thus, we end up with
\begin{subequations}
\begin{align}
\langle \hat{S}_z^2 \rangle_\mathrm{HFB} &= 0,
\\
\langle \hat{S}_z^2 \rangle_\mathrm{CPMFT} &= \mathrm{Tr}(\mathbf{K}^2).
\end{align}
\label{Sz}
\end{subequations}

The contribution to $\langle \hat{S}^2 \rangle$ from 
$\hat{S}_- \hat{S}_+$ must also be evaluated using the model two-particle 
density matrix.  Expanding this operator in terms of contributions from individual 
electrons, we have
\begin{equation}
\hat{S}_{-} \, \hat{S}_{+} = \underbrace{\sum_i \hat{s}_{-}(i) \hat{s}_{+}(i)}_{\hat{X}} + \underbrace{\sum_{i \ne j} \hat{s}_{-}(i) \hat{s}_{+}(j)}_{\hat{Y}}.
\end{equation}
The first term is the one-particle operator $\hat{X}$, and the second is the
two-particle operator $\hat{Y}$.

Since $\hat{X}$ does nothing to down-spin electrons but annihilates up-spin 
electrons, we clearly have
\begin{equation}
\langle \hat{X} \rangle = N_\beta = \mathrm{Tr}(\mathbf{P}).
\end{equation}
To take the expectation value of $\hat{Y}$, it proves useful to symmetrize it 
so that it acts the same on the two electrons.  Since operators acting on 
different electrons commute, we have
\begin{subequations}
\begin{align}
\hat{Y} &= \sum_{i \ne j} \hat{s}_{-}(i) \hat{s}_{+}(j)
\\
        &= \frac{1}{2} \sum_{i \ne j} \left( \hat{s}_{-}(i) \hat{s}_{+}(j) + \hat{s}_{+}(i) \hat{s}_{-}(j)\right)
\label{YSymm}
\\
        &= \sum_{i > j}  \left( \hat{s}_{-}(i) \hat{s}_{+}(j) + \hat{s}_{+}(i) \hat{s}_{-}(j)\right).
\end{align}
\end{subequations}
The only nonzero matrix elements of $\hat{Y}$ are
\begin{subequations}
\begin{align}
Y^{i_\beta j_\alpha}_{k_\alpha l_\beta} &= \langle i_\beta j_\alpha | \hat{Y} | k_\alpha l_\beta \rangle = \delta^i_k \delta^j_l,
\\
Y^{i_\alpha j_\beta}_{k_\beta l_\alpha} &= \langle i_\alpha j_\beta | \hat{Y} | k_\beta l_\alpha \rangle = \delta^i_k \delta^j_l.
\end{align}
\end{subequations}

The relevant spin components of the CPMFT and HFB model two-particle density 
matrix are
\begin{subequations}
\begin{align}
\Gamma_{i_\alpha j_\beta}^{k_\beta l_\alpha} &= \frac{1}{2}\left(-\gamma_{i_\alpha}^{l_\alpha} \gamma_{j_\beta}^{k_\beta} \mp \kappa_{i_\alpha j_\beta} \kappa^{k_\beta l_\alpha}\right),
\\
\Gamma_{i_\beta j_\alpha}^{k_\alpha l_\beta} &= \frac{1}{2}\left(-\gamma_{i_\beta}^{l_\beta} \gamma_{j_\alpha}^{k_\alpha} \mp \kappa_{i_\beta j_\alpha} \kappa^{k_\alpha l_\beta}\right),
\end{align}
\end{subequations}
where again CPMFT (HFB) corresponds to the top (bottom) sign.

Contracting the two-particle density matrix with the matrix elements gives us
\begin{equation}
\langle \hat{Y} \rangle = -\mathrm{Tr}(\bm{\gamma}_{\alpha\alpha} \bm{\gamma}_{\beta\beta} \mp \bm{\kappa}_{\alpha\beta} \bm{\kappa}_{\alpha\beta}).
\end{equation}
In the closed-shell case, using the results in Eqn. (\ref{ClosedShellStuff}), 
this becomes
\begin{equation}
\langle \hat{Y} \rangle = -\mathrm{Tr}(\mathbf{P}^2 \mp \mathbf{K}^2).
\end{equation}
Then the expectation value of $\hat{S}_- \hat{S}_+$ is given by
\begin{subequations}
\begin{align}
\langle \hat{S}_- \hat{S}_+ \rangle &= \mathrm{Tr}(\mathbf{P} - \mathbf{P}^2 \mp \mathbf{K}^2)
\\
                                    &= \mathrm{Tr}(\mathbf{K}^2 \mp \mathbf{K}^2).
\end{align}
\end{subequations}
We therefore have
\begin{subequations}
\begin{align}
\langle \hat{S}_- \hat{S}_+ \rangle_\mathrm{HFB} &= 2 \mathrm{Tr}(\mathbf{K}^2),
\\
\langle \hat{S}_- \hat{S}_+ \rangle_\mathrm{CPMFT} &= 0.
\end{align}
\label{SminSplus}
\end{subequations}

Combining Eqns. (\ref{Sz}) and (\ref{SminSplus}) gives us the total spin contamination in HFB and in CPMFT:
\begin{subequations}
\begin{align}
\langle \hat{S}^2 \rangle_\mathrm{HFB} &= 2 \mathrm{Tr}(\mathbf{K}^2),
\\
\langle \hat{S}^2 \rangle_\mathrm{CPMFT} &= \mathrm{Tr}(\mathbf{K}^2),
\end{align}
\end{subequations}

For UHF in cases in which there is strong correlation, we have the familiar 
formula
\begin{equation}
\langle \hat{S}^2 \rangle = s(s+1) + N_\beta - \mathrm{Tr}(\bm{\gamma}_{\alpha\alpha} \, \bm{\gamma}_{\beta\beta}).
\end{equation}
For the closed-shell case, using the results in Eqn. (\ref{ClosedShellUHF}), we have
\begin{align}
\langle \hat{S}^2 \rangle
 &= \mathrm{Tr}[\mathbf{P} - (\mathbf{P}+\mathbf{M})(\mathbf{P}-\mathbf{M})]
\\
 &= \mathrm{Tr}(\mathbf{P} - \mathbf{P}^2 + \mathbf{M}^2)
\nonumber
\\
 &= 2 \, \mathrm{Tr}(\mathbf{M}^2).
\nonumber
\end{align}


\begin{thebibliography}{18}
\bibitem{CPMFT1} T.~Tsuchimochi and G.~E.~Scuseria, J. Chem. Phys. {\bf 131}, 121102 (2009). 
\bibitem{CPMFT2} G.~E.~Scuseria and T.~Tsuchimochi, J. Chem. Phys. {\bf 131}, 164119 (2009).
\bibitem{CPMFT3} T.~Tsuchimochi, G.~E.~Scuseria, and A. Savin, J. Chem. Phys. {\bf 132}, 024111 (2010).
\bibitem{Blaizot} J-P.~Blaizot and G.~Ripka, ``{\it Quantum Theory of Finite Systems}" (The MIT Press, Massachusetts, 1986).
\bibitem{SS2002} V.~N.~Staroverov and G.~E.~Scuseria, J. Chem. Phys. {\bf 117}, 11107 (2002).
\bibitem{Yamaguchi} K.~Takatsuka, T.~Fueno, and K.~Yamaguchi, Theoret. Chim. Acta {\bf 48}, 175 (1978).
\bibitem{Star-David} V.~N.~Staroverov and E.~R.~Davidson, Chem. Phys. Lett. {\bf 330}, 161 (2000).
\bibitem{Mayer} I.~Mayer, Chem. Phys. Lett. {\bf 440}, 357 (2007).
\bibitem{Kutzelnigg} W.~Kutzelnigg and D.~Mukherjee, J. Chem. Phys. {\bf 110}, 2800 (1999). 
\bibitem{CHF} G. Cs${\rm {\acute a}}$nyi and T.~A.~Arias, Phys. Rev. B {\bf 61}, 7348 (2000).
\bibitem{Pople} J.~A.~Pople and R.~K.~Nesbet, J. Chem. Phys. {\bf 22}, 571 (1954).
\bibitem{FOOTNOTE} Note that UHF subtracts 
$\langle ij | kl \rangle M_{il} M_{jk}$ from the closed shell energy while 
CPMFT subtracts $\langle ij | kl \rangle  K_{ij} K_{kl}$.  However, these are 
the same if the basis functions and the anomalous density matrix are real, as 
they generally are.
\bibitem{Fukutome} H.~Fukutome, Int. J. Quantum Chem. {\bf 20}, 955 (1981).
\bibitem{Amos} A.~T.~Amos and G.~G.~Hall, Proc. Roy. Soc. (London) {\bf A263}, 483 (1961).
\bibitem{Harriman} J. E. Harriman, J. Chem. Phys. {\bf 40}, 2827 (1964).
\bibitem{math} V.~Rabanovich, Linear Algebra Appl. {\bf 390}, 137 (2004).
\bibitem{FOOTNOTE2} More precisely, the eigenvalues of 
$\alpha \, \mathbf{A} + \beta \, \mathbf{B}$ for idempotent $\mathbf{A}$ and 
$\mathbf{B}$ are 0, $\alpha$, $\beta$, or a corresponding pair 
$(n,\alpha+\beta-n)$.
\bibitem{GAUSSIAN} M.~J.~Frisch, G.~W.~Trucks, H.~B.~Schlegel, \textit{et al.}, Gaussian Development Version, Revion G.01, Gaussian, Inc., Wallingford CT, 2007.
\end{thebibliography}
\end{document}